\documentclass[epjST]{svjour}
\usepackage{mathptmx}
\usepackage{amsmath}
\usepackage{graphicx}
\usepackage{citesort}
\usepackage{color}
\usepackage{float}
\usepackage{placeins}

\usepackage[normalem]{ulem} 

\newcommand{\bd}{b^{\dag}}
\newcommand{\ban}{b^{\phantom{\dag}}}
\newcommand{\cd}{c^{\dag}}
\newcommand{\can}{c^{\phantom{\dag}}}
\newcommand{\dd}{d^{\dag}}
\newcommand{\dan}{d^{\phantom{\dag}}}

\begin{document}

\title{Quantifying the impact of phonon scattering on electrical and thermal
transport in quantum dots}

\author{B. Goldozian \thanks{\email{bahareh.goldozian@teorfys.lu.se}}
  \and G. Kir\v{s}anskas\and F.A. Damtie \and A. Wacker \thanks{\email{Andreas.Wacker@fysik.lu.se}} }
\institute{Mathematical Physics and NanoLund, Lund University, Box 118, 22100 Lund, Sweden}
\abstract{We report the inclusion of phonon scattering to our recently
established numerical package QmeQ for transport in quantum dot
systems. This enables straightforward calculations for a large variety
of devices. As examples we show (i) transport in a double-dot
structure, where energy relaxation is crucial to match the energy
difference between the levels, and (ii) the generation of electrical
power by contacting cold electric contacts with quantum dot states,
which are subjected to heated phonons.
} 
\headnote{\small Accepted Manuscript\\ to appear in European Physical Journal ST, special volume dedicated to FQMT17}
\maketitle
\section{Introduction}
\label{intro}
During the last decades, there has been a large amount of research
on the electronic properties and electron transport in nanostructure
devices \cite{DevoretNature1992,TaruchaPRL1996,KouwenhovenScience1997,ReimannRMP2002}.
A tremendous progress has been made in this area and new exciting optical,
electronic, and thermoelectric applications have been suggested \cite{ThelanderMaterialsToday2006,BoukaiNature2008,AgarwalApplPhysA2006,KoskiPNAS2014,KarimiNL2018}.

The confinement of electrons in nanostructure devices leads to discrete electronic states and hence to
quantized electron transport \cite{KastnerRMP1992,AverinPRB1991,FultonPRL1987,AshooriNature1996,KouwenhovenZPhysBCondMat1991}.
Therefore the transport properties become influenced by this quantization
that can be seen as Coulomb blockade \cite{AverinJLTP1986,AverinPRB1991,DevoretNature1992} and resonant
tunneling \cite{GurvitzPRB1996}. Likewise the electrons, phonons can also be confined in nanostructure devices.
The confinement of phonons leads to different modes with a one-dimensional
dispersion \cite{StroscioJAP1994}. An example of such effects is the temperature dependence of resistivity in Si
nanowires \cite{VauretteAPL2008}. Another  example is probing of the individual phonon modes in  transport \cite{WeberPRL2010}.
This new kind of phonon spectroscopy allows to get information on the energetic location of the phonon modes as well as
electron phonon interaction strength. 

To understand the role of different interactions in the many-body
physics of nanostructures, some studies have focused on the details of
electron-electron \cite{KuoPRB2010,AzemaPhysRevB2014}, electron-phonon
\cite{LeijnsePRB2010}, and electron-photon
\cite{HenrietPhysRevB2015} coupling effects on nonlinear
thermoelectric transport.  Moreover, a lot of progress has been made
both in formulating the many-body theories and developing experimental
methods in this subject. The  recent technological progress in the
design and fabrication of semiconductor nanostructures can help to
achieve a better understanding of how the electron-phonon coupling
affects the transport in nanostructures.

There is a large variety of theoretical approaches to calculate
transport through nanostructure
systems.\cite{BruusBook2004,TimmPRB2008,RyndykBook2009,SchallerBook2014}
For systems, where Coulomb interaction is dominant, single particle
transmission approaches are insufficient and different types of master
equations have been established. Recently, some of us published a
software package called QmeQ ({\bf Q}uantum {\bf m}aster {\bf
  e}quation for {\bf Q}uantum dot transport calculations)
\cite{KirsanskasComputPhysCommun2017}, which allows a simple
comparison of several Master equations for quantum dot systems with
full Coulomb interaction. Here we show the first results of phonon
scattering integration into this package.

As a model system, we consider a spinful double dot with tunable
energy levels and compare with experimental data from
\cite{WeberPRL2010}. Double quantum dots are well-known to be
suitable systems to study essential concepts of quantum mechanics, due
to their high tunability.  Furthermore, we consider non-equilibrium
phonon distributions, which can establish particle current flow
against the bias \cite{SothmannNJP2013}.

 \section{QmeQ and its extension for phonon interaction}
 \label{sec:theory}
 The package QmeQ \cite{KirsanskasComputPhysCommun2017} is designed to study transport in quantum dot systems described by the general
 Hamiltonian,
\begin{equation}
H_\textrm{sys} = \underset{n\sigma}{{\displaystyle \sum}}E_{n}\dd_{n\sigma}\dan_{n\sigma}+\underset{nm\sigma}
{{\displaystyle \sum}}\Omega_{nm}\dd_{n\sigma}\dan_{m\sigma}+ \frac{1}{2}\underset{mnkl}{{\displaystyle \sum}}
\underset{\sigma \sigma^{\prime}}
{{\displaystyle \sum}}
V_{mnkl}\dd_{m\sigma^{\phantom{\prime}}}\dd_{n\sigma^{\prime}}\dan_{k\sigma^{\prime}}\dan_{l\sigma},\label{eq:HamiltonianD}
\end{equation}
where $\dd_{n\sigma}$ is the electron creation operator in
the system for a level with energy $E_{n}$ and spin $\sigma$. $\Omega_{nm}$ describes the coupling between the
single-particle levels  $m$ and $n$, and
the last part quantifies the electron-electron interaction with different Coulomb matrix elements $V_{mnkl}$.
Using a Fock basis, $H_\textrm{sys}$ is diagonalized providing the many-particle eigenstates of the form
\begin{equation}
\left|a\right\rangle =\underset{n}{\sum}b_{n}\dd_{n_1}\dd_{n_2}...\dd_{n_{Na}}\left|0\right\rangle,
\end{equation}
where $n_{i}$ represents the $i$th single particle state in the Slater determinant determined by the index $n=(n_{1},n_{2,},...,n_{Na})$ with $N_{a}$ number of particles.

The tunneling and the lead Hamiltonians are
\begin{equation}
H_{\textrm{Leads}}=\underset{k\sigma l}{{\displaystyle \sum}}E_{k\sigma \ell}\cd_{k\sigma l}\can_{k\sigma \ell},\quad
H_{T} = \underset{n,k\sigma \ell}{{\displaystyle \sum}}(t_{n\ell}\dd_{n\sigma}\can_{k\sigma \ell}+\mathrm{H.c.}),
\end{equation}
where $\cd_{k\sigma \ell}$
denotes the electron creation operators in the leads with index $\ell$ (typically R,L for right and left lead,
respectively) and  $k$ labels the spatial wave-functions of the continuum of states. We  assume a thermal distribution
with electrochemical potential $\mu_\ell$ and temperature $T_\ell$ for each lead.
A bias $V$ applied to the leads provides $\mu_{L/R}=\pm eV/2$, where $e$ is the elementary charge.
The tunneling  Hamiltonian, $H_{T}$,
can be expressed by the many-particle eigenstates
\begin{equation}
H_{T}=\underset{ab,k\ell\sigma}{\sum}(T_{_{ba}}(\ell\sigma)\left|b\right\rangle \left\langle a\right|\can_{k\ell\sigma}+\mathrm{H.c.})
\quad \textrm{with} \quad T_{_{ba}}(\ell\sigma)=\underset{n}{\sum}t_{n\ell}\left\langle b\right|\dd_{n\sigma}\left|a\right\rangle .
\end{equation}
Here we use the letter convention restricting the combination of states $|a\rangle,|a\prime\rangle, |b\rangle$ to $N_{b}=N_{a}+1$, and $N_{a}=N_{a\prime}$.

QmeQ is able to calculate the current for a stationary state
with different approaches: the Pauli (classical) master equation
\cite{BreuerBook2006,BruusBook2004},
first-order Redfield \cite{WangsnessPR1953,RedfieldIBM1957}
and first/second-order von Neumann approaches \cite{PedersenPRB2005a,PedersenPHE2010},
and a master equation in a Lindblad form \cite{LindbladCMP1976} using the Position and Energy Resolved Lindblad Approach (PERLind) of Ref.~\cite{KirsanskasPRB2018}.

In this study we calculate the impact of electron-phonon interaction on the particle current through the device with
different first-order approaches. The phonons are modeled as simple non-interacting Bosonic modes
\begin{equation}
H_{ph}=\underset{q}{\sum}\hbar\omega_{q}\bd_{q}\ban_{q},
\end{equation}
were $\bd_{q}$ creates a phonon in a mode $q$. The electron-phonon interaction is given by
\begin{equation}
H_{e-ph} = \underset{nm\sigma,q}{\sum}g_{nm}^{q}\dd_{n\sigma}\dan_{m\sigma}(\bd_{\overline{q}}+\ban_{q})\label{eq:phononH}.
\end{equation}
with the matrix elements $g_{nm}^{q}$ and $\overline{q}$ denoting the
complex conjugate state of $q$.  Since the phonon coupling to the free
electrons in the leads is very small compared to the electron phonon
coupling in the central region, see Ref.~\cite{WingreenPRB1989}, only
the electron phonon coupling in the dot will be considered. Within the
lowest nonvanishing order in the phonon coupling studied here,
coherent superpositions  of states with different phonon number are
neglected.

We consider deformation potential coupling to the phonons, which is given
by the divergence of the displacement following \cite{LindwallPRL2007}.
The corresponding coupling matrix element for the first acoustic phonon mode coupled
to the electrons via the deformation potential can be expressed by

\begin{equation}
g_{nm}^{q1}=\int d^{3}r\, \varPsi_{n}^{\ast}(\mathbf{r})D\nabla\cdot\mathbf{u}_{q1}(\mathbf{r})\varPsi_{m}(\mathbf{r}).\label{eq:gphonon}
\end{equation}
Here $\mathbf{u}_{q1}(\mathbf{r})$ is the displacement and $D$ is the deformation potential coefficient. We express the 
electron-phonon coupling matrix element, $g_{nm}^{q}=g(q)y_{nm}^{q}$, in terms of a state-independent overall strength $g(q)$ and a dimensionless coefficient
\begin{equation}
y_{nm}^{q}=\int d^{3}r\, \varPsi_{n}^{\ast}(\mathbf{r})\textrm{e}^{\textrm{i} \mathbf{q}\cdot \mathbf{r}}(\mathbf{r})\varPsi_{n^{\prime}}(\mathbf{r})\label{eq:gphonon2}.
\end{equation}
By assuming that $y_{nm}^{q}$ is $q$-independent (e.g. by choosing a characteristic value), we obtain
\begin{equation}
y_{nm}^{q}\approx y_{nm},\quad
y_{nm}^{\overline{q}}\approx \overline{y}_{nm}=y_{mn}^{*},
\end{equation}
and we  can collect all the energy dependence in the spectral density
\begin{equation}
J(E)=\underset{q}{\sum}\left|g(q)\right|^{2}\delta(E-\hbar\omega_{q}).
\end{equation}
Both $J(E)$ and $y_{nm}$
are required inputs in the new version of QmeQ with the possibility to add further vibrational modes as independent  processes in an analogous way.

\section{Results}
\label{sec:Results1}
In order to illustrate the performance of QmeQ, we simulate the
nanowire double-dot system studied  in Ref.~\cite{WeberPRL2010}, and
compare the results with experimental data. Each dot has a ground level
$E_{L/R}$ and an excited level $E_{L/R}+5.5$ meV (both spin degenerate), where
$E_L$ and $E_R$ can be individually  shifted by plunger gates.
The two dots are coupled
to each other and to the leads by tunnel barriers.
We use the InAs nanowire material parameters (see
Ref.~\cite{WeberPRL2010}) to calculate the spectral density for the lowest,
one-dimensional phonon mode:

\begin{equation}
J(E)=3.8804\times10^{-4}E.
\end{equation}

The matrix elements $y_{nm}$ for the relevant single-particle
levels $m$ and $n$ are calculated from Eq. (\ref{eq:gphonon2}), where
the corresponding electron wave functions are assumed to be
$\varPsi_{1}(\mathbf{r})=a^{-{3}/{2}}\pi^{-{3}/{4}}e^{-{(x^{2}+y^2+z^{2})}/{2a^{2}}}$
for the ground state and $\varPsi_{3}(\mathbf{r})=\sqrt{2}za^{-{5}/{2}}\pi^{-{3}/{4}}e^{-{(x^{2}+y^2+z^{2})}/{2a^{2}}}$
for the excited state of the left dot with
the Gaussian radius $a=5.8$ nm \cite{LindwallPRL2007}.
The corresponding states for the right dot ($n=2,4$) are shifted by $d=120$ m, in $z$-direction. For $qa\ll 1$,
$q$ only enters as the phase  $e^{iqd}$ for matrix elements with different dots, and we use a typical value
$e^{iqd}=e^{i\pi/3}$ to remove the $q$-dependence.

The electron-electron Coulomb
matrix elements are calculated with the same methods as described in
\cite{GoldozianSciRep2016}.  We consider intradot interactions, $U=12$
meV, as the interaction between electrons in the same dot and
inter-dot interaction, $U_{n}=2.5$ meV, as the interaction between
electrons in the neighboring dots.  The  couplings between the energy
levels in the dots and the left and right leads are ${\Gamma_L}$ = 90
neV and ${\Gamma_R}$ = 10 neV, respectively \cite{WeberPRL2010}.
 \begin{figure}[t]
 \begin{center}
 \resizebox{0.85\columnwidth}{!}{%
  \includegraphics{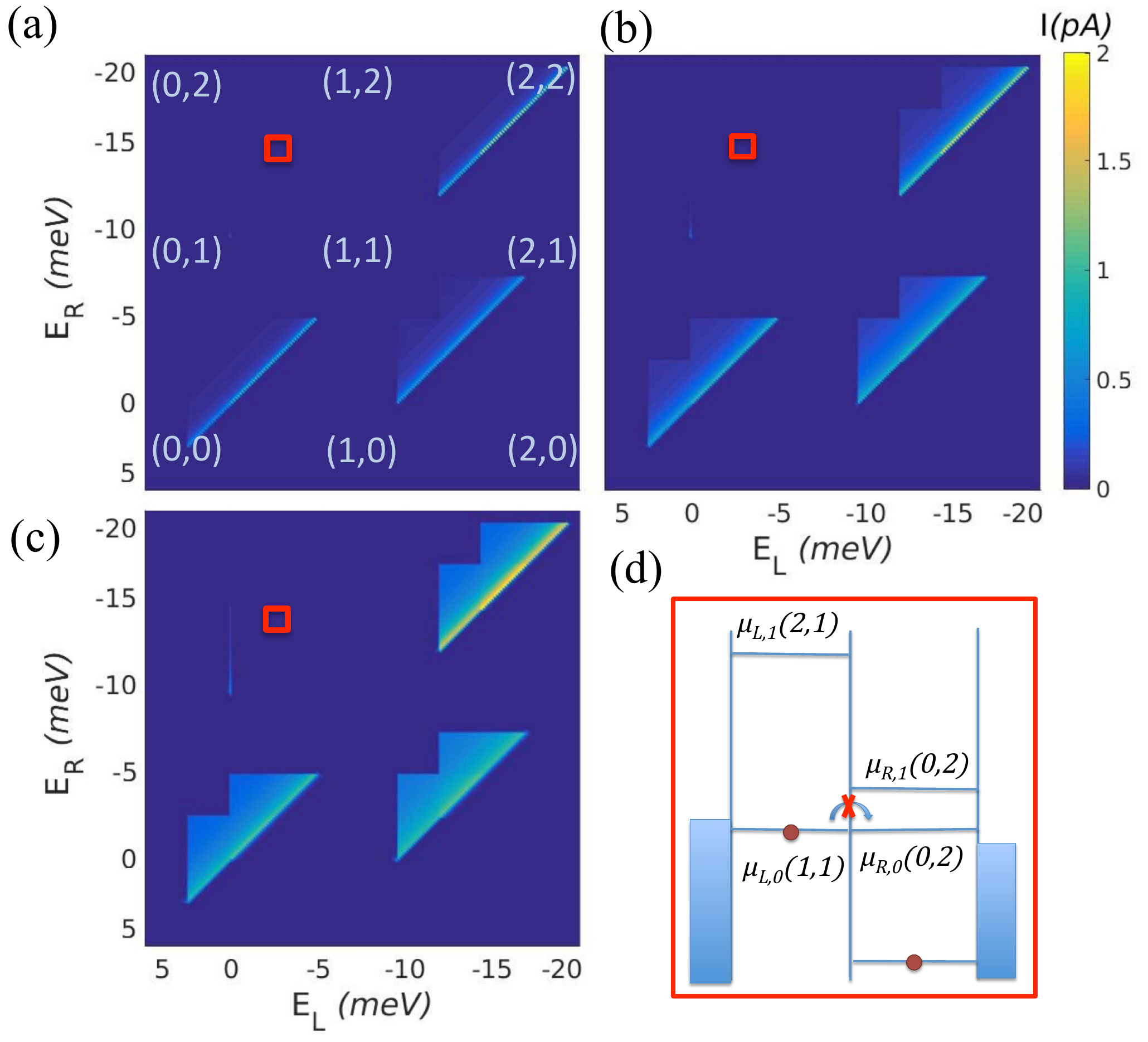} }
  \end{center}
 \caption{Stability diagrams of the double dot system ranging from weak to strong inter-dot tunnel coupling, i.e.,
(a) $\Omega$ = 0.005 meV, (b) $\Omega$ = 0.01 meV (c) $\Omega$ = 0.02 meV. The bias is $V=5$ mV  for all
 of them. The diagram in (d) shows schematically the removal energies $\mu_{i,n}(N_{l},N_{r})$
 at the operation point denoted by red squares in (a), (b), and (c). At this point the current is suppressed due to Pauli spin blockade.
[$\mu_{i,n}(N_{l},N_{r})$ is the energy of an electron removed from the level $n$ ($0=$ ground, $1=$ excited)
of dot $i$ (left or right) where $N_{l/r}$ is the number of electrons in the left/right dot.] }
 \label{fig:1}
 \end{figure}

Fig.~\ref{fig:1} shows the current as a function of the left
and the right dot energy levels ($E_{L}$ and $E_{R}$) for three different tunnel
coupling strengths $\Omega$.
We can see the lowest inter-dot transitions between
different level configurations of the double dot, which give rise
to the current.
What stands out in this figure is the current increase in the presence
of phonon scattering close to the current peaks
from the resonance. This is a result of the electron-phonon
scattering which is an inelastic process. The triangles of finite current appear due to the phonon
emission process. When the energy level in the left dot is higher than the one in the right dot, electrons in the left dot
are able to transport through the right dot by emitting phonons. The absorption of phonons does not play any role
since $k_BT\ll eV$.
We observed a pair of overlapping full bias triangles, for each transition,
due to the inter-dot Coulomb interactions. These triangles are more pronounced in the plots corresponding to stronger
tunnel coupling between the dots (Fig.~\ref{fig:1}(b) and Fig.~\ref{fig:1}(c)).

On the left top of the plots current suppression due to Pauli spin blockade \cite{OnoScience2002,JohnsonPRB2005} can be seen.
This corresponds
to the transition  $(1,1)\longrightarrow (0,2)$, where current is blocked, if both dots are occupied by the same spin.
Fig.~\ref{fig:1}(d) shows
the energy levels at this operation point.

 \begin{figure}[t]
 \begin{center}
 \resizebox{0.95\columnwidth}{!}{%
  \includegraphics{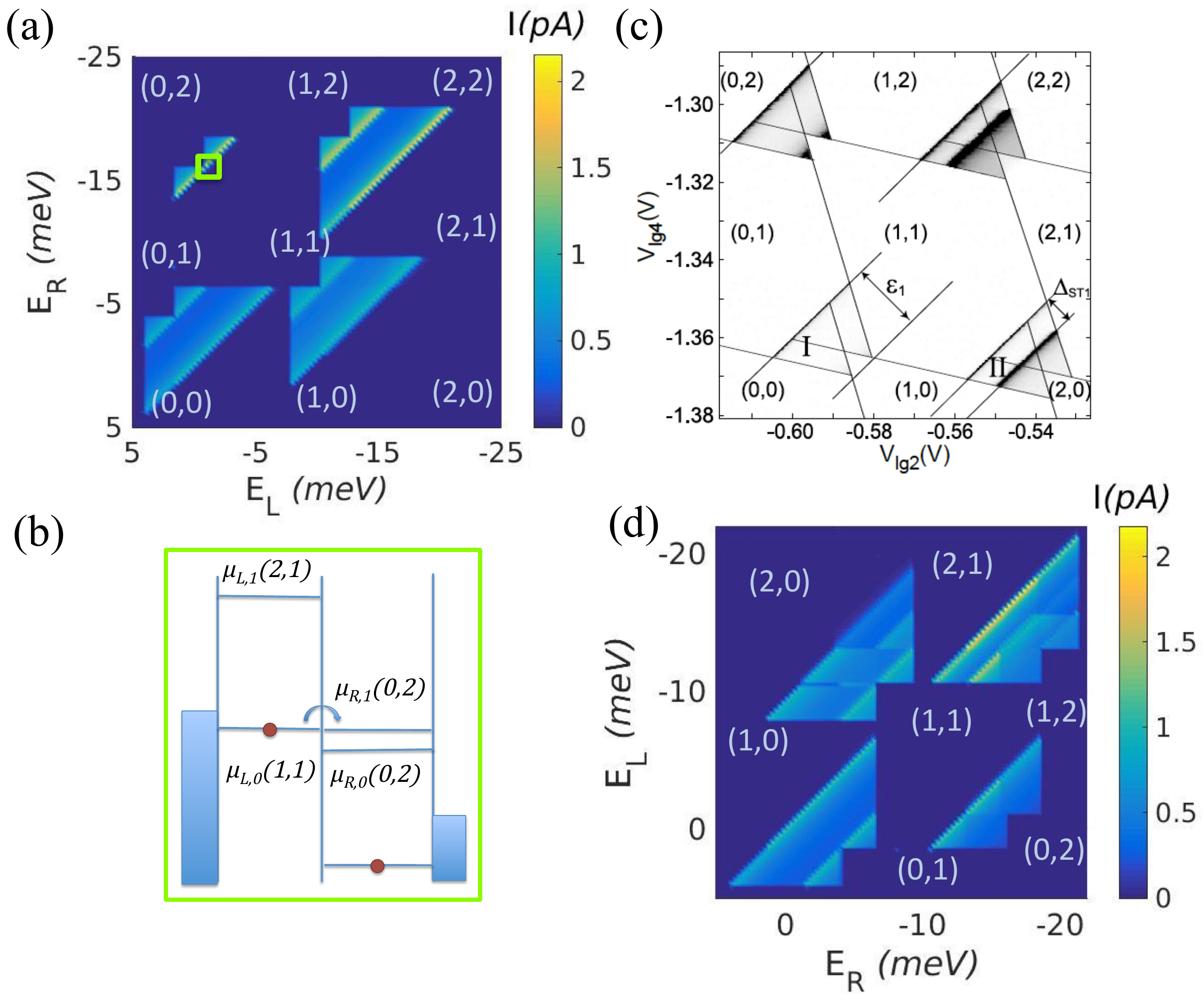} }
  \end{center}
 \caption{(a) Stability diagram of the double dot system with the inter-dot tunnel coupling
$\Omega$ = $0.02$ meV and an increased bias $V=8$ mV compared to Fig.~\ref{fig:1}. Here the combination of electron-phonon
coupling and high bias lifts the Pauli spin blockade as sketched in panel (b). (c) Experimental data (with different bias polarity) 
taken from the supplementary material of \cite{WeberPRL2010}. The highest current is $\approx 1$ pA. (d) Stability diagram for the same parameters as panel (a)
including the exchange interaction, $U_{ex}=3$ meV, displayed with swapped axes to reflect the different bias polarity in (c).}
 \label{fig:2}
 \end{figure}

Figure \ref{fig:2}(a) depicts the results for the same system as
Fig.~\ref{fig:1}, but with higher bias.  As we can see here, in
contrast to Fig.~\ref{fig:1}, the transition  $(1,1)\longrightarrow
(0,2)$ is possible, since the high bias creates a new transport
channel via the excited state. As the schematic diagram in
Fig.~\ref{fig:2}(b) explains, electrons form the excitation
$\mu_{L,0}(1,1)$ can now be in resonance with  $\mu_{R,1}(0,2)$, while
the lower lying two-particle state  $\mu_{R,0}(0,2)$ is still emptied
to the right lead.  The triangular regions as well as the truncated
triangle between the (1,1) and (0,2) region fully agree with
experimental data displayed  in  Fig.~\ref{fig:2}(c). Note that the
bias polarity was chosen differently in the experiment, so that (0,2)
compares to (2,0) in our definition.  While the line due to tunneling
via the excited state is shifted from the ground state by a fixed
amount 5.5 meV for all triangles in our simulation, see
Fig.~\ref{fig:2}(a), two different separations can be seen in the
experiment, depending on the charging of the receiving quantum
dot. This difference can be attributed to the exchange interaction
within the dots. Assuming $U_{ex}=3$ meV, Fig.~\ref{fig:2}(d) provides
a better agreement with the experiment.  We observe, that
Fig.~\ref{fig:2}(d) contains some additional features in the triangles
with two electrons in the left dot, if a spin triplet becomes possible.
These are not observed in the experiment, most likely due to spin
relaxations on nanosecond (or shorter) time scales.  Such
spin-relaxation also explains the incomplete spin blockade observed in
the experiment.

\begin{figure}[t]
\begin{center}
 \resizebox{0.95\columnwidth}{!}{%
  \includegraphics{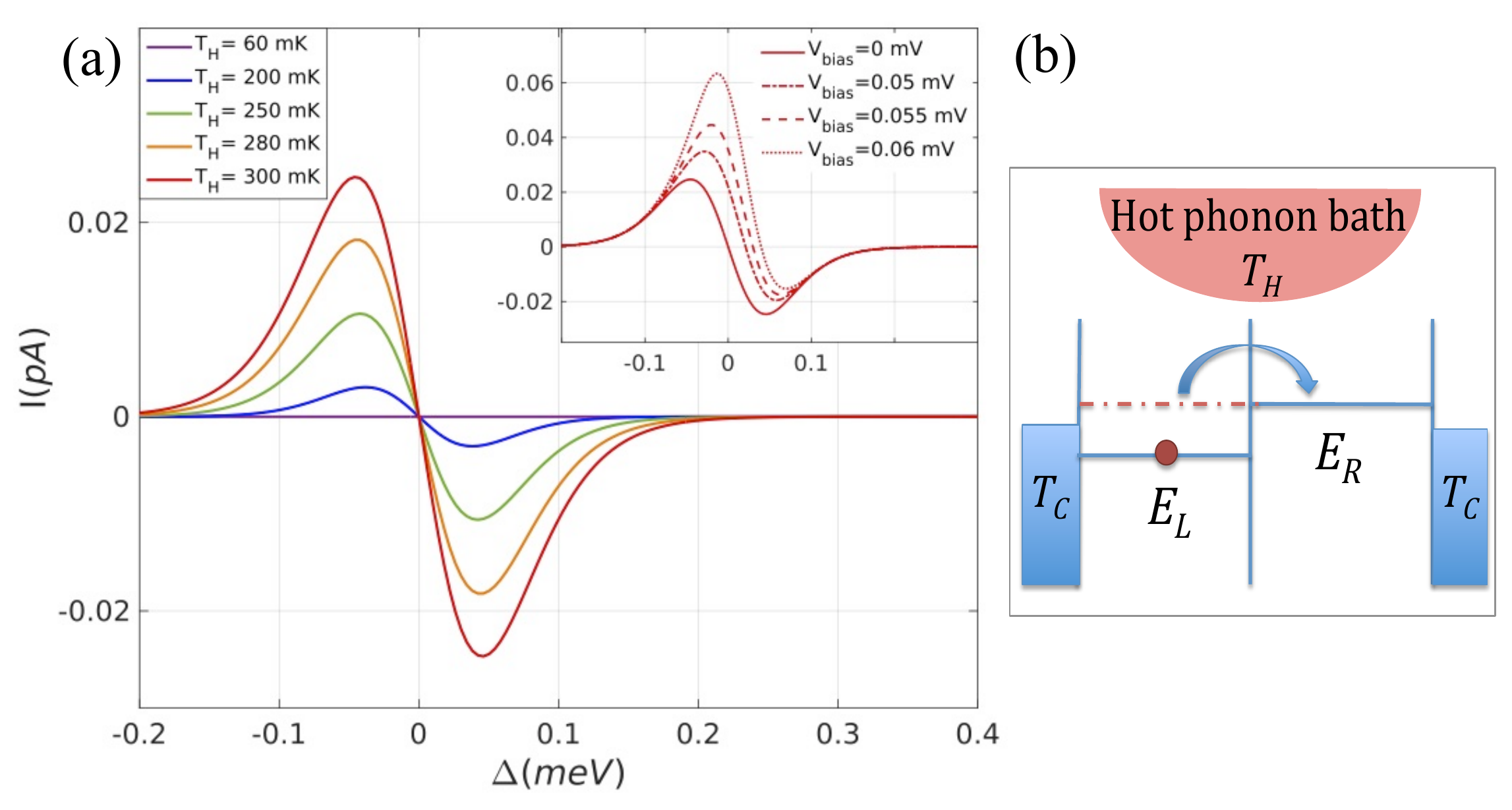} }
  \end{center}
 \caption{(a) Zero-bias current as a function of $\Delta$, the difference between the energy levels of the dots,
   for different phonon temperatures. Inset: current as a function of $\Delta$ for different biases and $T_{H} = 300$ mK.
   Further parameter: $T_{C}$ = 60 mK, $\Omega$ = 0.05 meV, ${\Gamma_L}$ = ${\Gamma_R}$ = 90 neV.
 (b) A sketch of a double dot system coupled to non-equilibrium phonon bath.}
 \label{fig:3}       
 \end{figure}

As a second example we consider the impact of a heated phonon
distribution in the double dot system, which can act as a heat engine
\cite{SothmannNJP2013,JiangPhysRevApplied2017}. A similar scenario can
also arise due to a noise source\cite{EntinPRB2017}. Here we consider
the same system as before, but neglect the excited levels.  The setup
is sketched in the right panel of Fig.~\ref{fig:3}.  While keeping the
electron leads at $T_C=60$ mK, we apply a different temperature $T_H$
in the phonon distribution.  Fig.~\ref{fig:3} shows the current as a
function of the difference between the energy levels of the dots,
$\Delta = E_{L}$ - $E_{R}$, which are shifted symmetrically with
respect to the electrochemical potential in the leads at zero bias.
We find, that an asymmetry in the level energies drives a current
through the quantum dot system, where (for $T_H>T_C$)  the net
particle current goes from the lower to the higher level. The reason
is that the thermal distributions in the contacts result  in a
significant difference between the occupations of the upper and lower
dot level. For the larger phonon temperature, this implies  a
dominance of phonon absorption over phonon emission between these
levels, actually driving the current by taking heat from the phonon
system. As thoroughly discussed in
Refs.~\cite{SothmannNJP2013,JiangPhysRevApplied2017} this acts as a
heat engine, if this current flows against an electric bias. This can
be seen in the inset:  For a positive detuning  $\Delta \approx 0.08$
meV, i.e., $E_L>E_R$, the current flows against the bias polarity.  In
contrast to the treatment of Ref.~\cite{JiangPhysRevApplied2017} the
use of QmeQ allows for a straightforward  implementation of the full
many-body interaction in the transport calculations.

\begin{figure}[t]
\begin{center}
 \resizebox{1\columnwidth}{!}{%
  \includegraphics{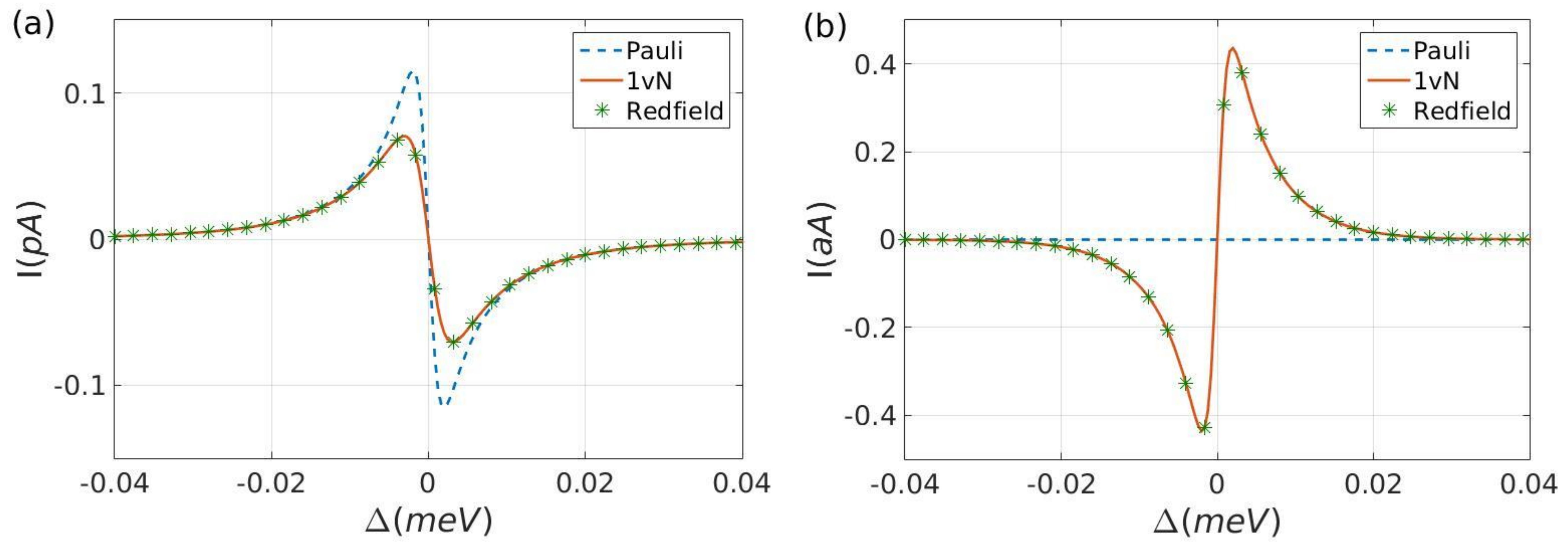} }
  \end{center}
 \caption{(a) Zero-bias current as a function of $\Delta$, the difference between the energy levels of the dots,
   for Pauli, Redfield and 1vN approaches. $T_{C}$ = 60 mK, $T_{H}$ = 300 mK,  $\Omega$ = 0.001 meV, ${\Gamma_L}$ = ${\Gamma_R}$ = 0.005 meV.
 (b) Zero-bias current as a function of $\Delta$, $T_{C}$ = $T_{H}$ = 60 mK, other parameters are the same as (a).}
 \label{fig:4}       
 \end{figure}

These calculations were performed with the Pauli master equation,
while the Redfield and  first order von Neumann (1vN) approaches gave
the same results for these parameters, where ${\Gamma}$ is much
smaller than the level splitting and non-diagonal elements of the
density matrix are not relevant
\cite{KirsanskasComputPhysCommun2017,GoldozianSciRep2016}.  However,
in the case where $\Omega$ is less than ${\Gamma}$, as it is shown in
Fig.~\ref{fig:4}, coherences become important and the Pauli master
equation is not reliable.  Fig.~\ref{fig:4}(a) shows the current as a
function of $\Delta$, for Pauli, Redfield, and 1vN approaches.  As it
can be seen the current is reduced in the Redfield and 1vN approaches
where the coherences are important. Fig.~\ref{fig:4}(b) displays the
zero bias current as a function of $\Delta$ for  $T_{C}$ = $T_{H}$ =
60 mK. Here no current should flow which is recovered by the Pauli
master equation. However, the Redfield and 1vN approaches show a very
small current as they do not fully satisfy thermal detailed
balance. These violations are actually three orders of magnitude
smaller than the features studied in Fig.~\ref{fig:4}(a) and thus do
not affect the main results.

 \section{Conclusion}
\label{sec:Conclusion}
We included phonon scattering in QmeQ, which improves the applicability of
this versatile simulation package. The new implementation was tested for a double-dot
structure, where we could reproduce experimental results and demonstrated
 the applicability for thermoelectric elements.
\\[0.2cm]
{\bf Acknowledgments:} We thank NanoLund and the Swedish Research Council (Grant No. 621-2012-4024) for financial support. 
\\[0.2cm]
{\bf Author Contributions:} BG performed the calculations, produced the figures, and wrote the first draft. GK established the computer program using input from an earlier code by FAD. AW directed the work. All authors contributed to the writing of the manuscript.

\FloatBarrier

\end{document}